\documentclass[preprint2]{aastex}

\slugcomment{Accepted for publication in ApJL}

\shorttitle{WASP-4b: Southern transiting planet}
\shortauthors{Wilson et al.}

\begin{document}

\title{WASP-4b: a 12th-magnitude transiting hot-Jupiter in the Southern hemisphere}

\author{
D.M. Wilson\altaffilmark{1},
M. Gillon\altaffilmark{2,12},
C. Hellier\altaffilmark{1},
P.F.L. Maxted\altaffilmark{1},
F. Pepe\altaffilmark{2},
D. Queloz\altaffilmark{2},
D.R. Anderson\altaffilmark{1},
A. Collier Cameron\altaffilmark{3},
B. Smalley\altaffilmark{1},
T.A. Lister\altaffilmark{1,7},
S.J. Bentley\altaffilmark{1},
A. Blecha\altaffilmark{2},
D.J. Christian\altaffilmark{4},
B. Enoch\altaffilmark{5},
C.A. Haswell\altaffilmark{5},
L. Hebb\altaffilmark{3},
K. Horne\altaffilmark{3},
J. Irwin\altaffilmark{6,11},
Y.C. Joshi\altaffilmark{4},
S.R. Kane\altaffilmark{14},
M. Marmier\altaffilmark{2},
M. Mayor\altaffilmark{2},
N. Parley\altaffilmark{5},
D. Pollacco\altaffilmark{4},
F. Pont\altaffilmark{2},
R. Ryans\altaffilmark{4},
D. Segransan\altaffilmark{2},
I. Skillen\altaffilmark{8},
R.A. Street\altaffilmark{4,7,13},
S. Udry\altaffilmark{2},
R.G. West\altaffilmark{9},
P.J. Wheatley\altaffilmark{10}
}

\altaffiltext{1}{Astrophysics Group, Keele University, Staffordshire, ST5 5BG, UK} 
\altaffiltext{2}{Observatoire de Gen\`{e}ve, 51 ch. des Maillettes, 1290 Sauverny, Switzerland} 
\altaffiltext{3}{School of Physics and Astronomy, University of St. Andrews, North Haugh,  Fife, KY16 9SS, UK} 
\altaffiltext{4}{Astrophysics Research Centre, School of Mathematics \& Physics, Queen's University, University Road, Belfast, BT7 1NN, UK} 
\altaffiltext{5}{Department of Physics and Astronomy, The Open University, Milton Keynes, MK7 6AA, UK} 
\altaffiltext{6}{Institute of Astronomy, University of Cambridge, Madingley Road, Cambridge, CB3 0HA, UK} 
\altaffiltext{7}{Las Cumbres Observatory, 6740 Cortona Dr. Suite 102, Santa Barbara, CA 93117, USA} 
\altaffiltext{8}{Isaac Newton Group of Telescopes, Apartado de Correos 321, E-38700 Santa Cruz de la Palma, Tenerife, Spain} \altaffiltext{9}{Department of Physics and Astronomy, University of Leicester, Leicester, LE1 7RH, UK} 
\altaffiltext{10}{Department of Physics, University of Warwick, Coventry, CV4 7AL, UK}
\altaffiltext{11}{Harvard-Smithsonian Center for Astrophysics, 60 Garden Street MS-16, Cambridge, MA 02138-1516, USA}
\altaffiltext{12}{Institut d'Astrophysique et de G\'eophysique,  Universit\'e de Li\`ege, 4000 Li\`ege, Belgium}
\altaffiltext{13}{Department of Physics, Broida Hall, University of California, Santa Barbara, CA 93106-9530, USA}
\altaffiltext{14}{Michelson Science Center, Caltech, MS 100-22, 770 South Wilson Avenue, Pasadena, CA  91125, USA}

\begin{abstract}
\sloppy
We report the discovery of WASP-4b, a large transiting gas-giant planet with an orbital period of 1.34 days.  This is the first planet to be discovered by the SuperWASP-South observatory and CORALIE collaboration and the first planet orbiting a star brighter than 16$^{th}$ magnitude to be discovered in the Southern hemisphere.  A simultaneous fit to high-quality lightcurves and precision radial-velocity measurements leads to a planetary mass of 1.22$^{+0.09}_{-0.08}$ M$\rm_{Jup}$ and a planetary radius of 1.42$^{+0.07}_{-0.04}$ R$\rm_{Jup}$. The host star is USNO-B1.0 0479-0948995, a G7V star of visual magnitude 12.5.
As a result of the short orbital period, the predicted surface temperature of the planet is 1761 K, making it an ideal candidate for detections of the secondary eclipse at infrared wavelengths.
\end{abstract}


\section{Introduction}
Transiting extrasolar planets offer our best opportunity for measuring fundamental parameters of extrasolar planets, such as radius and mass, allowing us to test our models for planetary formation and evolution.
To-date the only transiting planets known in the Southern sky are those discovered by surveys targeting the galactic plane (e.g. \citealp{ogle, ogle2,lupus}) and find planets around stars with typical visual magnitudes of 16--17. 

The SuperWASP-South (WASP-S) instrument looks for transiting planets around stars of visual magnitude 8--13, and will eventually cover the entire southern sky, excluding the Galactic plane.  We report here the discovery of WASP-4b, a transiting hot-Jupiter orbiting a star of magnitude 12.5.

This is the first discovery
by WASP-S, made in collaboration with radial-velocity measurements from the CORALIE spectrograph on the 1.2-m Euler telescope. The discovery marks the beginning of a campaign to discover brighter transiting extrasolar planets in the Southern hemisphere, to complement those discovered in the North by projects including HAT \citep{hat}, TrES \citep{tres}, XO \citep{xo} and WASP \citep{sw}.

Transit searches are most sensitive to large planets with short orbital periods, and have now found nine Jupiter-mass planets in orbits of less than 2 d\footnote{http://exoplanet.eu}. This has led to suggestions of a class of very-hot Jupiters \citep{melo}, with highly irradiated atmospheres (e.g. \citealt{fortney}), to which WASP-4b likely belongs.

\section{Photometric Observations}

The WASP (Wide Angle Search for Planets) consortium operates two identical robotic observatories; SuperWASP-North on La Palma, in the Northern hemisphere, and SuperWASP-South, in the Southern hemisphere, hosted by the South African Astronomical Observatory (SAAO).  Each consists of eight wide-field cameras consisting of an 11.1-cm aperture Canon 200 mm f/1.8 lens backed by a 2k$\times$2k e2v CCD.  The eight cameras cover 490 square degrees per pointing.
WASP-S started operating in May 2006, with a strategy of tiling six to eight fields with a cadence of 5--10 mins and exposure times of 30 secs.  These fields rotate with the seasons, accumulating to a strip centered on a declination of --32$^\circ$, resulting in lightcurves for over one million stars brighter than 13$^{th}$ magnitude.   For details of the WASP project,
hardware and data processing see \citet{sw}.  Further details of our data analysis procedures are given in \citet{wasp1}, reporting the discovery of  WASP-1b and WASP-2b from SuperWASP-North, and in \citet{wasp3} reporting the discovery of WASP-3b.

\begin{figure}[h]
\centering
\includegraphics[angle=0,width=0.45\textwidth]{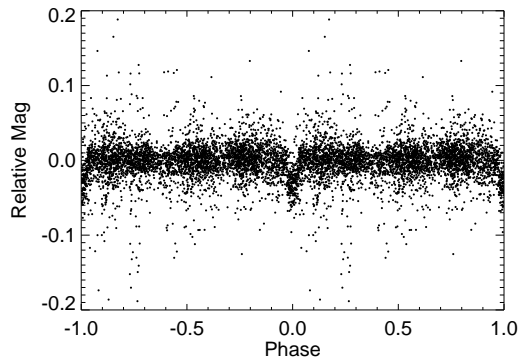}
\caption{The SuperWASP-South discovery lightcurve of WASP-4 folded on the 1.3-d orbital period.}
\label{figure:sw}
\end{figure}

From the WASP-S data collected between 2006 May--November we identified 1SWASP\,J233415.06--420341.1 (USNO-B1.0 0479-0948995) as a high-priority planetary transit candidate; over 4000 measurements revealed a possible transit recurring every 1.3 days (see \citealt{hunter} for a description of our transit-search methods and \citealt{mcmc} for an account of our selection of high-priority candidates).  A catalogue search within the 48"
photometric aperture of 1SWASP\,J233415.06--420341.1 revealed no bright blending companions or known variable/active stars, which could mimic the photometric signature of a transiting planet.  The WASP-S discovery lightcurve is shown in Figure~\ref{figure:sw}.

\begin{figure}[h]
\centering
\includegraphics[angle=0,width=0.40\textwidth]{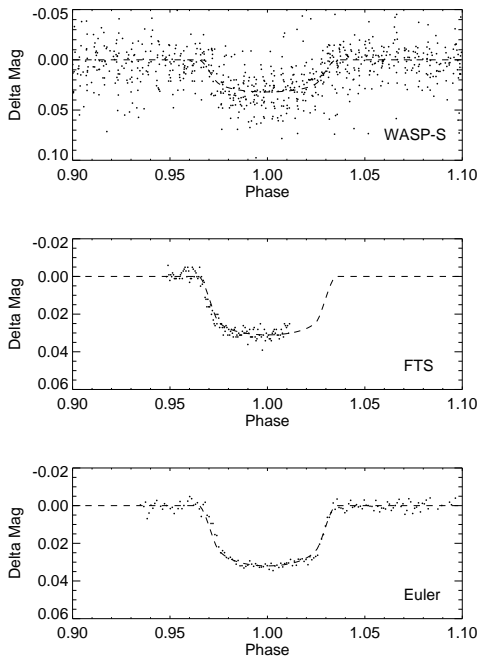}
\caption{The WASP-S, FTS and EulerCam lightcurves showing the transit of WASP-4b. The data are shown folded on the orbital period together with the best-fitting model determined from a simultaneous MCMC fit of the photometric and radial-velocity data. The RMS scatter to the model fit of the WASP-S, FTS and EulerCam lightcurves is 15.3, 2.7 and 1.8 mmags respectively.  Phase 1 for the EulerCam lightcurve occurred at HJD=2454368.59121 UT and for the FTS lightcurve at HJD=2454371.26766 UT.}
\label{figure:phot} 
\end{figure}

A full transit of WASP-4b was observed by EulerCAM on the 1.2-m Euler telescope on 2007 September 25$^{th}$. Observations were performed in R-band and were heavily de-focussed to allow exposure times of $\sim$2 mins, achieving an RMS scatter of 1.8 mmags despite poor transparency conditions. Part of a transit was also observed in SDSS i' band by the 2.0-m Faulkes Telescope South (FTS) on 2007 September 27$^{th}$. The WASP-S, EulerCAM and FTS lightcurves are shown in Figure~\ref{figure:phot}.

\section{Spectroscopic Observations with \\CORALIE} 
\begin{figure}[h]
\centering
\includegraphics[angle=0,width=0.45\textwidth]{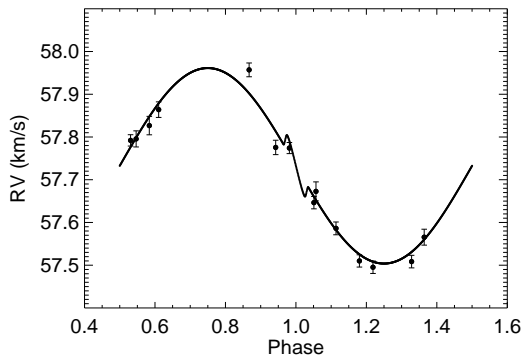}
\caption{CORALIE radial-velocity measurements folded on the orbital period together with a radial-velocity model which includes the expected Rossiter--McLaughlin effect for a star with $v \sin i$ = 2 km\,s$^{-1}$. The RMS of the RV measurements to the model fit is 15.3 m\,s$^{-1}$.}
\label{figure:rv}
\end{figure}

Spectroscopic measurements were obtained using the CORALIE spectrograph installed on the Euler telescope.  CORALIE was originally a twin copy of the ELODIE spectrograph \citep{elodie}. However, triggered by the interest to carry out spectroscopic follow-up on transit candidates with this instrument, in June 2007 we carried out major changes to CORALIE to increase its performance on fainter stars. The fibre link and the cross-disperser optics have been removed and replaced by a new design (Queloz et al. in prep.). The double scrambler has also been removed and the grism and prism cross-disperser component replaced by a series of 4 Schott F2 prisms of 32$^\circ$ angle each.  The net outcome of this new design is to maintain the spectral range from 381 to 681 nm but with a large efficiency gain of about a factor of 6 (8 below 420 nm) and a spectroscopic resolution of 55\,000--60\,000 (increased by 10--20\%). The overall instrumental precision is also improved.  All exposures are reduced by the automated pipeline adapted to the new optical configuration. WASP-4 was observed from 2007 September to November and the radial velocity and the $v \sin i$ computed using a G2-spectral template. Exposures were 30 minutes in length, achieving a signal-to-noise ratio of $\sim$15. Radial velocity variations of semi-amplitude 251 m\,s$^{-1}$ were detected consistent with a planetary-mass companion whose orbital period closely matches that from the WASP-S transit detections.  The RV measurements are listed in Table~\ref{table:rv} and are shown phase folded and over plotted with the best-fitting orbital model in Figure~\ref{figure:rv}. The RMS of the RV measurements to the model fit is 15.3 m\,s$^{-1}$.

\begin{table}[h]
\centering
\caption{CORALIE RV measurements for WASP-4.}
\label{table:rv}
\begin{tabular}{ccc} 
\tableline\tableline 
BJD -- 2450\,000 & RV (km\,s$^{-1}$) & $\sigma _{RV}$ (km\,s$^{-1}$)\\ 
\tableline
\\
4359.71082 &	57.56557 &	0.01864 \\
4362.63121 &	57.79551 &	0.01883 \\
4364.65260 & 	57.67271 &	0.02223 \\
4365.73690 &	57.95703 &	0.01613 \\
4372.75799 &	57.58619 &	0.01496 \\
4376.68882 & 	57.64642 &	0.01451 \\
4378.66887 &	57.79207 &	0.01325 \\
4379.73630 &	57.50846 &	0.01470 \\
4380.61034 &	57.77411 &	0.01303 \\
4382.79025 &	57.86396 &	0.01823 \\
4383.55277 &	57.51000 &	0.01434 \\
4387.61904	&	57.49524 &	0.01457 \\
4408.66110	&	57.77556 &	0.01682 \\	
4409.51932	&	57.82666 &	0.02156 \\
\\
\tableline\tableline
\end{tabular}
\end{table}

\subsection{Line-bisector analysis}
\begin{figure}[h]
\centering
\includegraphics[angle=0,width=0.45\textwidth]{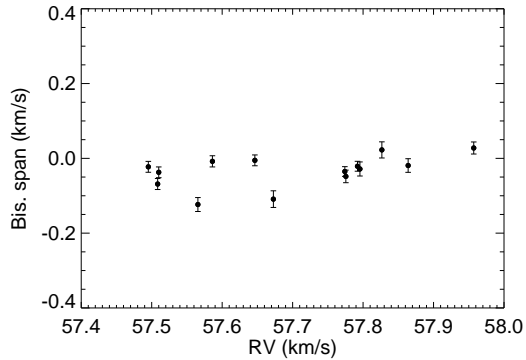}
\caption{Line bisector spans for the CORALIE radial velocity measurements showing no significant variation.}
\label{figure:bis_spans}
\end{figure}

Although the amplitude of the RV variation of this system is consistent with a planetary companion, the signal could be mimicked by spectral line distortions caused by a blended eclipsing binary (e.g. \citealp{santos}).  To confirm the presence of a transiting planet it is necessary to exclude this scenario.  Blended eclipsing binaries can be identified from multi-colour photometry or through lightcurve modeling \citep{torres}, however, the simplest and most reliable method is to search for variations in the spectral line profiles themselves.  This has the added advantage of highlighting atmospheric distortion effects which can also affect the measured radial velocities.  If the detected radial velocity variations are the result of contamination of the spectrum by an eclipsing binary, we would expect to see distortions in the line profiles in phase with the photometric period.  The CORALIE cross-correlation functions were analysed using the line-bisector technique \citep{bisector} and no evidence for variation in the bisector spans was found, confirming the planetary nature of this system. A plot of the bisector spans is shown in Figure~\ref{figure:bis_spans}.

\section{Stellar parameters}
The individual CORALIE spectra have a low signal-to-noise ratio, but when co-added (and re-binned) give a S/N of around 30--40 which is suitable for a preliminary photospheric analysis of WASP-4. The analysis was performed using the {\sc uclsyn} spectral synthesis package and {\sc atlas9} models without convective overshooting \citep{castelli}. The H$\alpha$, Na {\sc i} D and Mg {\sc i} b lines were used as diagnostics of both $T_{\rm eff}$ and $\log g$. The metallicity was estimated using the photospheric lines in the 6000--6200 \AA\ region. However, the co-added spectrum is not of sufficient quality to perform a detailed abundance analysis.  The parameters obtained from this analysis are listed in Table~\ref{table:stellar_params}.  In addition to the spectral analysis, we have also used Tycho B, V and 2MASS magnitudes to estimate the effective temperature using the Infrared Flux Method \citep{blackwell}. This gives $T_{\rm eff} = 5410 \pm 240$~K, which is in close agreement with that obtained from the spectroscopic analysis. These results imply a spectral type of around G7.

In our spectra the Li {\sc i} 6708 \AA\ line is not detected, allowing us to derive an upper limit on the Lithium abundance of log n(Li/H) +
12 $<$ 1.0, which is slightly less than the Solar photospheric value.
This implies a minimum age of around 2 Gyr \citep{sestito}. Comparison of the temperature and $\log g$ with the stellar evolution models of \citet{girardi} gives maximum-likelihood values $M_* = 0.90 \pm 0.07$ M$_\sun$ and $R_* = 1.15 \pm 0.28$ R$_\sun$. The distance of WASP-4 was calculated as 300 $\pm$ 50 pc using the angular diameter from the Infrared Flux Method and the value of the radius of the star from Table~\ref{table:mcmc_params}.

\begin{table}[h]
\centering
\caption{Stellar parameters of WASP-4 derived from a spectral analysis of the CORALIE data.}
\label{table:stellar_params}
\begin{tabular}{lr}
\tableline\tableline
Parameter & Value\\
\tableline
\\
R.A. (J2000.0) & 23$\rm^{h}$34$\rm^{m}$15.06$\rm^{s}$ \\ 
Dec. (J2000.0) & --42$^\circ$03$^{'}$41.1$^{''}$ \\ \\ 
$T_{\rm eff}$ (K) & 5500 $\pm$ 150 \\ 
log ${\it g}$ & 4.3 $\pm$ 0.2 \\ 
$[$M/H$]$  & 0.0 $\pm$ 0.2 \\ 
$\log n({\rm Li})$  & $<$ 1.0 \\
$v \sin i$ (km\,s$^{-1}$) & 2.2 $^{+0.6}_{-1.0}$ \\\\ 
Mass (M$_{\sun}$) &  0.90 $\pm$ 0.07 \\ 
Radius (R$_{\sun}$) &  1.15 $\pm$ 0.28 \\ 
Spectral Type & G7V \\
Distance (pc) & 300 $\pm$ 50\\
\\
\tableline\tableline \end{tabular} \end{table}

\section{Planetary parameters}
The photometric and orbital parameters for WASP-4b were determined from a simultaneous Markov-Chain Monte-Carlo fit of both the photometric and spectroscopic data.  The MCMC method is described in detail in \citet{mcmc} and is the same used for the fitting of the parameters of WASP-3b \citep{wasp3}.

The transit lightcurve was modelled using the small-planet approximation of \citet{mandelagol}.  An initial fit of the radial velocity measurements showed no evidence for significant orbital eccentricity, as expected for such a short-period system. Accordingly the eccentricity was set to zero.
This leaves seven parameters defining the system: epoch ($T_{0}$), orbital period ($P$), duration of the transit ($t_{T}$) from first to last contact, the squared ratio of the planet to stellar radius ($\Delta F = (R_{P}/R{\star})^{2}$), the impact parameter ($b = a \cos i/R\star$), the radial velocity amplitude ($K_{1}$) and the stellar mass $M_{\star}$.  Gaussian priors were imposed, as determined from the spectroscopic analysis, such that $M_{\star}$ = 0.90 $\pm$ 0.07 and log ${\it g}$ = 4.3 $\pm$ 0.2.  
To balance the weight of the radial velocities with the photometry in the MCMC analysis we added 7 m\,s$^{-1}$ of systematic error to the errors listed in Table~\ref{table:rv}, thus obtaining a reduced Chi-squared of 1.

The resulting fitted parameters are listed, together with their 1-$\sigma$ uncertainties, in Table~\ref{table:mcmc_params}.  The parameters derived for the host star are consistent with the spectral analysis from the CORALIE data. The precision of the transit photometry is sufficient to constrain the stellar radius and impact parameter such that the uncertainties in the resultant planetary parameters are dominated by the uncertainty in the stellar mass.

\begin{table}[h]
\caption{Fitted system parameters of WASP-4b from a simultaneous MCMC analysis of the WASP-S, FTS and EulerCam lightcurves together with the CORALIE RV data, assuming a circular orbit.}
\label{table:mcmc_params}
\begin{tabular}{lll}
\tableline\tableline
Parameter & Value & Error\\
\tableline
\\
Period (days) & 1.3382282 & $^{+0.000003}_{-0.000003}$ \\ 
Epoch (HJD) & 2454365.91464 & $^{+0.00025}_{-0.00023}$ \\ 
Duration (days) & 0.0928 & $^{+0.0009}_{-0.0007}$ \\ 
($R\rm_{P}$/$R_{\star}$)$^{2}$ & 0.0241 & $^{+0.0005}_{-0.0002}$ \\ 
$b$  & 0.13 & $^{+0.13}_{-0.12}$ \\ 
$i$ (degrees)  & 88.59 & $^{+1.36}_{-1.50}$ \\

\\

$K\rm_{1}$ (km\,s$^{-1}$) & 0.24 & $^{+0.01}_{-0.01}$ \\  
$\gamma$ (km\,s$^{-1}$) & 57.7326 & $^{+0.002}_{-0.001}$ \\   
$a$ (AU) & 0.0230 & $^{+0.001}_{-0.001}$ \\ 
log ${\it g}$ (cgs) & 4.45 & $^{+0.016}_{-0.029}$ \\ 
$R_{\star}$ (R$_{\sun}$) &  0.9370 & $^{+0.04}_{-0.03}$ \\ 
$M_{\star}$ (M$_{\sun}$) &  0.8997 & $^{+0.077}_{-0.072}$ \\ 
$\rho_{\star}$ ($\rho_{\sun}$) &  1.094 & $^{+0.038}_{-0.085}$ \\ 

\\

$R\rm_{P}$/R$_{\rm Jup}$ &  1.416 & $^{+0.068}_{-0.043}$ \\  
$M\rm_{P}$/M$_{\rm Jup}$ &  1.215 & $^{+0.087}_{-0.079}$ \\  
$\rho_{P}$ ($\rho_{\rm Jup}$) &  0.428 & $^{+0.032}_{-0.044}$ \\ 
log $g_{P}$ (cgs) & 3.142 & $^{+0.023}_{-0.034}$ \\  
T$_{P}$ (A=0)(K) &  1761 & $^{+24}_{-9}$ \\
\\
\tableline\tableline
\end{tabular}
\end{table}

\section{Discussion}

A simultaneous fit of precision photometry and radial velocity measurements result in a planetary radius of 1.42 R$_{\rm Jup}$ and mass of 1.22 M$_{\rm Jup}$ for WASP-4b.  This is the second largest transiting planet discovered to date, second only to TrES-4b (\citealp{tres4}; for plots comparing WASP-4b with other transiting extrasolar planets see \citealt{wasp3}).

WASP-4b, with an orbital period of 1.3 d, can be compared to other very-short-period planets such as OGLE-TR-56b \citep{ogle56}, TrES-3b \citep{tres3} and WASP-3b \citep{wasp3}.  The low rotation velocity of the host star WASP-4 is comparable with that of TrES-3; in contrast, OGLE-TR-56 and WASP-3 have much higher rotation velocities, which is thought to be due to the young age of these stars.  The stars  WASP-4 and TrES-3 are thought to be older, however the synchronisation time for WASP-4 would be longer still at 8 Gyr (using the  method of \citealp{marcy}), so their low rotational velocity is consistent with their not having been spun-up by their planets.

A preliminary estimate of the stellar parameters indicates that the parent star is spectral type G7V with solar metallicity. Together with the short orbital period of 1.34 d this results in a blackbody planetary surface temperature, assuming zero albedo and isotropic re-radiation, of 1760 K.  The orbital distance of 0.023 AU places WASP-4b well within the criterion ($a < 0.04$ AU) for the proposed new class of pM planets \citep{fortney}.  These planets display a temperature inversion due to low-pressure stellar absorption by gaseous TiO and VO.  As a result they exhibit an unusually hot stratosphere which emits strongly in the mid-infrared.  The WASP-4b system has a larger flux ratio than the very-hot-Jupiters WASP-3b and TrES-3b and so is an ideal target for secondary eclipse studies by {\sl Spitzer}.  The study of more of these systems, including WASP-4b, should help constrain models for atmospheric dynamics and heat distribution.

\acknowledgments
\section{Acknowledgments}

The WASP Consortium includes the Universities of Keele, Leicester, St.~Andrews, Queen's University Belfast, The Open University and the Isaac Newton Group.  WASP-S is hosted by the South African Astronomical Observatory (SAAO) and we are grateful for their support and assistance. Funding for WASP comes from Consortium Universities and the UK's Science and Technology Facilities Council.

\end{document}